\documentclass[a4paper]{aa}
\input{psfig}
\def\eg{{\it e.g.} }
\def\etal{{\it et al.} }
\def\ie{{\it i.e. } }
\def\dt{\frac{\partial}{\partial t}}
\def\dk{\frac{\partial}{\partial k}}
\def\tdt{\frac{\tilde\partial}{\partial t}}

\def\dd #1 {{\frac{\partial}{\partial #1}}}

\def\cs2{c_{S}^2}

\def\ltsima{$\; \buildrel < \over \sim \;$}
\def\simlt{\lower.5ex\hbox{\ltsima}}
\def\gtsima{$\;\buildrel>\over\sim\;$}
\def\simgt{\lower.5ex\hbox{\gtsima}}
 
\authorrunning{Tagger}
\titlerunning{Rossby waves and vortices}
\begin{document}
\title{On Rossby waves and vortices with differential rotation}
\author{M.Tagger\inst{1}}
\offprints{M.~Tagger (tagger@cea.fr)}
\institute{DSM/DAPNIA/Service d'Astrophysique (CNRS URA 2052), CEA 
Saclay, 91191 Gif-sur-Yvette, France}
\date{Received date; accepted date}
\abstract{
We present a simplified model for linearized perturbations in
a fluid with both differential rotation and differential vorticity. 
Without the latter the model reduces to the classical Shearing Sheet
used in the description of spiral density waves in astrophysical disks. 
Without the former it reduces to the $\beta$-plane approximation, used
in the description of Rossby waves.  Retaining both, our model allows
one in general to discuss the coupling between density waves and Rossby
waves, resulting in what is known as the ``corotation resonance'' for
density waves.  Here we will derive, as a first application of this
model, the properties of Rossby waves in a differentially rotating disk. 
We find that their propagation is quenched by differential rotation:
after a limited number of oscillations, a Rossby wave collapses to a
singular vortex, as fluid elements are sheared apart by differential
rotation.  In a keplerian disk, this number of oscillations is always
lower than one.  We also describe how, in a similar manner, a vortex is
sheared in a very short time.
  }
\maketitle

\section{Introduction}
\label{sec:intro}
Spiral density waves are well known in differentially rotating disks;
they can be propagated by pressure alone, producing the
Papaloizou-Pringle instability (Papaloizou and Pringle, 1984); or they
can rely on long-range forces: self gravity, as in galactic disks or
Saturn's rings (see {\em e.g.} Binney and Tremaine, 1987, and references
therein), or magnetic stresses (Tagger {\em et al.}, 1990; Tagger and
Pellat, 1999).  Their nature is that of compressional waves, \ie sound
waves modified by differential rotation and long-range forces. They can
be excited by various interactions (in particular tidal forces), or
linearly unstable.

On the other hand Rossby waves are widely studied in the context of
planetary atmospheres, with solid rotation (associated with the planet)
but differential vorticity, due to the latitudinal gradient.  Their
nature is essentially that of a torsional wave, propagating vorticity. 
Their basic properties are derived in a convenient model, the
``$\beta$-plane approximation'', where one neglects the gradients of all
equilibrium quantities (including geometrical ones) except vorticity:
the physics is thus reduced to the minimal description needed to
describe Rossby waves.

Symmetrically, density waves are described in the ``Shearing Sheet''
model, where the sole gradient considered is differential rotation, and
in particular the gradient of vorticity is neglected.  These models are
very convenient since they allow one to consider the essential
properties of the waves, including their amplification.  Global models,
on the other hand, give more complete results but their complexity in
general precludes a detailed understanding of the basic physics
involved.  However the Shearing Sheet as well as the $\beta$-plane
approximation neglect quantities which may convey important physical
information.  In particular, in a disk, the
vorticity\footnote{hereafter, without any risk of ambiguity, we will
consider only what is actually the vertical component of vorticity} is:
\begin{equation}
    W=\frac{\kappa^2}{2\Omega}
    \label{eq:vorticity}
\end{equation}
where $\Omega$ is the rotation frequency, and $\kappa$ the epicyclic 
frequency given by:
\begin{equation}
    \kappa^2=4\Omega^2+2\Omega\Omega'\ r
    \label{eq:epicyclic}
\end{equation}
where the prime notes the radial derivative.  Thus in a keplerian disk
($\Omega\sim r^{-3/2}$) one has $W=\Omega/2$, so that its gradient is
comparable to differential rotation and should not {\em a priori} be
neglected.  Symmetrically, in planetary atmospheres, zonal circulation
is often fast enough that differential rotation might play an important
role in the propagation of Rossby waves.

In this paper we show how, in the cylindrical geometry of a disk, the
derivation of the Shearing Sheet model can be made more exact, while
retaining its essential simplicity.  This results in a set of equations
with two parameters, the gradients of rotation frequency and of
vorticity.  The model reduces to the $\beta$-plane approximation or to
the classical Shearing Sheet (CSS) respectively, when either parameter
vanishes.  In the former case one recovers Rossby waves, and in the
latter case usual spiral density waves.  In the general case where both
gradients are retained, the model allows one to describe how these
waves, loosing their identity because of differential rotation, become
coupled: density waves generating Rossby waves as they propagate, and
reciprocally Rossby waves spawning density waves as they are sheared by
differential rotation.  The corresponding exchange of energy and angular
momentum between the waves affects the linear stability of the disk, as
discussed below.  For this reason we call our model, appropriate to
describe this exchange of energy and momentum, the Rossby Shearing
Sheet.

The details and consequences of this coupling will be presented in
separate publications.  Here we will discuss only the basic properties
of the model, in section~\ref{sec:model}, and present in
section~\ref{sec:rossby} a first application: we will give an exact
solution in the incompressible limit, describing the propagation of
Rossby waves in presence of differential rotation.  We find that after a
finite number of oscillations a Rossby wave always collapses to a
singular vortex, as fluid elements participating in the wave motion are
sheared apart by differential motion.  We also discuss in
section~\ref{sec:vortex} how differential rotation shears a vortex in a
very short time, as found in recent numerical simulations.

The last section will be devoted to a short discussion of these results, 
their consequences and their connection with already existing ones.

\section{The Rossby Shearing Sheet}
\label{sec:model}
Our goal is to derive in a rigorous manner a form of the Shearing Sheet
model, that would allow us to take into account the vorticity gradient,
and thus the physics of Rossby waves.  In practice this will result in a
set of equations going smoothly from the classical Shearing Sheet (CSS,
no vorticity gradient) to the $\beta$-plane approximation (no
differential rotation).

We thus start from the linearized Euler and continuity equations, written
in cylindrical geometry:
\begin{eqnarray}
    \tdt v_{r}  -2\Omega v_{\vartheta}  =f_{r}
     \label{eq:euleru}\\
    \tdt v_{\vartheta}  +W v_{r}  = f_{\vartheta}
     \label{eq:eulerv}\\
    \tdt \rho  =-\rho_{0}\vec\nabla\cdot\vec V
     \label{eq:contin}
\end{eqnarray}

where $f=F/\rho_{0}$, $\rho_{0}$ is the equilibrium density, and the
force $F$ is left unspecified; we have assumed, in the spirit of the
CSS, a disk with constant density and temperature.  We have
defined: \[\tdt = \dt+im\Omega\] where $m$ is the azimuthal wavenumber. 

We use logarithmic coordinates with \[s=\ln r\] and define
\begin{eqnarray*}
    u & = & r v_{r}\\
    v & = & r v_{\vartheta}\\
    D & = & \frac{\partial}{\partial s}u+imv=r^2\vec\nabla\cdot\vec V\\
    R & = & \frac{\partial}{\partial s}v-imu=r^2\left(\vec\nabla\times\vec 
    V\right)_{z}\\
\end{eqnarray*}
Thus $D$ and $R$ represent the compressional and vortical components of 
the velocity field.

Combining equations (\ref{eq:euleru}-\ref{eq:contin}) we get:
\begin{eqnarray}
     \tdt D  -2\Omega R- 2\Omega'(v-imu)&=& r^2\vec\nabla\cdot\vec f%\hfill
    \label{eq:eulD}  \\
     \tdt R +WD+W'u &=&  r^2\vec\nabla\times\vec f%\hfill
    \label{eq:eulR}  \\
    \tdt h &=& - \frac{D}{ r^2}%\hfill
    \label{eq:conth}
\end{eqnarray}
where the $'$ now defines the derivative with respect to $s$, and
$h=\rho/\rho_{0}$.  These equations are exact in cylindrical geometry. 
In MHD disks threaded by a vertical magnetic field (see Tagger \& Pellat
1999) the term in $\vec\nabla\times\vec f$ gives access to the physics
of Alfven waves.  Here we restrict ourselves to conventional fluid disks
with only pressure stresses, so that this term disappears (it would also
disappear if we retained self-gravity), while\\
\[\vec\nabla\cdot\vec 
f=-\frac{c_{S}^2}{r^2}\left(\frac{\partial^2}{\partial 
s^2}-m^2\right)h \]
where $c_{S}$ is the sound speed, and we have assumed for simplicity an
isothermal equation of state.  By writing the equations for $D$ and $R$,
we have explicitly displayed the term in $W'$ in equation
(\ref{eq:eulR}), which does not appear in the usual derivation of the
shearing sheet.  Thus we can now use the same techniques as in the CSS,
considering as constant all the coefficients in the left-hand sides
except for the rotation frequency, linearized around the fiducial radius
$s=0$ (which can conveniently be taken as the corotation radius $r_{0}$,
if one treats a monochromatic wave).  Thus we write:
\begin{equation}
    \tdt=\dt+im(\Omega_{0}+2Ams)
    \label{eq:tom}
\end{equation}
where $\Omega_{0}=\Omega(s=0)$, and $A=\Omega'/2$ is Oort's constant. 
We perform a Fourier transform in $s$ (note that then the radial
wavenumber $k$ is non-dimensional), so that the term proportional to $s$
gives a derivative in $k$.  We work in the frame rotating at the
frequency $\Omega_{0}$, so that finally the time derivative becomes
\begin{equation}
    \tdt\longrightarrow\dt-2Am\dk
    \label{eq:dtdk}
\end{equation}
We use the {\it exact} relations:
\[u=\frac{-ikD+imR}{q^2}, \ v=\frac{-imD-ikR}{q^2}\]
where 
\[q^2=k^2+m^2\]
and get a third order differential system:
\begin{eqnarray}
    \left(\dt-2Am\dk\right)D& = &2\Omega
    R\nonumber\\
    +\displaystyle\frac{4A}{q^2} [-(k+i)&m&D+(m^2-ik)R]+q^2c_{S}^2h
    \label{eq:fourD}  \\
    \left(\dt-2Am\dk\right) R &=& 
    -WD+i\frac{\alpha}{q^2}\left(kD-mR\right)
    \label{eq:fourR}  \\
    \left(\dt-2Am\dk\right) h &=& -\frac{1}{r_{0}^2}D
    \label{eq:fourh}\\
\end{eqnarray}
where we have defined \[\alpha=W'\] (note that $\alpha=A$ in a keplerian
disk).  In the CSS, the last term disappears from the right-hand side of
equation (\ref{eq:fourR}), allowing us to reduce the system to second
order with
\begin{equation}
    h=Wr_{0}^2 R
    \label{eq:spcvort}
\end{equation}
This describes the Papaloizou and Pringle (1984, 1985) instability; with
additional forces it would describe self-gravity driven or magnetically
driven spiral instabilities.  Note that equation (\ref{eq:spcvort}) is
the linearized form of the more general conservation of specific
vorticity (also called vortensity in this context):
\[(\vec\nabla\times\vec V)/\rho\]
but that a model involving equilibrium radial density gradients would be
much less tractable than the present one; thus for simplicity we will
stick (in the spirit of the CSS and $\beta$-plane approximation) to the
description of a flow with no equilibrium gradients besides rotation and
vorticity.  For comparison with more complete models however, we will
need to keep in mind this difference.

The third order system must usually be solved numerically: the coupling
between the second-order system and the third equation, proportional to
$\alpha$, describes the exchange of energy and angular momentum between
the density and Rossby waves.  However we can get interesting results
already in the incompressible limit: setting $h=D=0$, the system reduces
to:
\begin{equation}
    \left(\dt-2Am\dk\right) R=-im\frac{\alpha}{q^2}R
    \label{eq:incomp}
\end{equation}

Let us first consider solid rotation with differential vorticity ($A=0,\ 
\alpha\neq 0$): equation (\ref{eq:incomp}) gives the dispersion relation of 
Rossby waves, in the $\beta$-plane approximation:
\begin{equation}
    \omega=\frac{m\alpha}{k^2+m^2}\label{eq:ross}
\end{equation}

Going back to the full compressible equations, and retaining both $A$
and $\alpha$, we use the little-known, but powerful formalism of Lin
and Thurstans (1984): we note that $t$ appears in equations
(\ref{eq:fourD}-\ref{eq:fourh}) only through the operator in the
left-hand sides.  We can thus separate variables, writing:
\begin{equation}
    \Phi(k,t)=\hat\Phi(k)F(k-2Amt)
\end{equation}
where $\Phi$ is the vector $(D,\ R,\ h)$.  Then the function $F(k-2Amt)$
can be factored out and disappears from the equations; it is an envelope
function, given by the initial conditions.  Two choices of $F$ have an
important physical meaning: $F=Cte$ corresponds to normal modes
(standing wave patterns).  This can easily be seen by writing the
inverse transform:
\begin{equation}
    \Phi(s,t)=\frac{1}{2i\pi}\int_{-\infty}^{+\infty} dk\ e^{iks}
    \ \hat\Phi(k)\ F(k-2Amt)
\end{equation}
so that, with $F=Cte$, $\Phi$ does not depend on $t$, in the frame
rotating at the frequency $\Omega_{0}$.  This corresponds to a standing
wave pattern, \ie a normal mode, for the discrete set of wave
frequencies (corresponding to a choice of $r_{0}$) such that the
appropriate boundary conditions are fulfilled.

In the same manner, one finds that $F=\delta(k-2amt-k_{0})$ corresponds
to a single wave, launched with $k=k_{0}$ at $t=0$.  This allows one to
study the transient fate of a single wave propagating in the disk, as
differential rotation causes its radial wavenumber $k$ (selected by the
delta function in $F$) to evolve with time, so that the wave pattern
changes from tightly wound leading (with $k$ large and negative) to open
($k\simeq 0$) to tightly wound trailing ($k\rightarrow +\infty$).

For simplicity we will neglect the $\hat{\ }$ below.  Equations
(\ref{eq:fourD}-\ref{eq:fourh}) become, for any choice of $F$:
\begin{eqnarray}
    -2Am\dk D & = & 2\Omega 
    R\nonumber\\
    +\frac{4A}{q^2} [-(k&+&i)mD+(m^2-ik)R]+q^2c_{S}^2h
    \label{eq:fourlinD}  \\
    -2Am\dk R & = & -WD+i\frac{\alpha}{q^2}\left(kD-mR\right)
    \label{eq:fourlinR}  \\
    -2Am\dk h & = & -\frac{1}{r_{0}^2}D
    \label{eq:fourlinh}
\end{eqnarray}
In equation (\ref{eq:fourlinR}), the term proportional to $\alpha$ now
describes an exchange of specific vorticity with the equilibrium flow. 
With $\alpha=0$, the system can be reduced to a second order ODE, in
which self-gravity and magnetic stresses are easily incorporated.  This
system has thus been the basis to analyze the amplification of spiral
waves driven by self-gravity, by magnetic stresses, or by pressure
forces.

As discussed above, on the other hand, in the incompressible limit
equation (\ref{eq:fourlinR}) represents the propagation of Rossby waves. 
In general the term proportional to $\alpha$ in equation
(\ref{eq:fourlinR}) thus represents a source of Rossby waves, generated
by the propagation of density waves. The Rossby wave exchanges energy 
and angular momentum with the spiral, thus causing amplification or 
damping.

The dependence of this mechanism on the vorticity gradient, and the
singularity of Rossby waves which will be discussed in the next section,
allow us to identify it with the ``corotation resonance'', known to
occur for density waves: this has been discussed in particular by
Panatoni (1983) for self-gravity driven spirals, by Papaloizou and
Pringle (1985), Papaloizou and Lin (1989) and Narayan \etal (1987) for
the Papaloizou-Pringle instability, and more recently by Tagger and
Pellat (1999) in the case of magnetically driven spirals.  In the first
two cases it was found to make only a minor difference on the properties
and growth rate of the instabilities.

In the latter case on the other hand it was found both quite efficient
(as compared with the weak instability found in the CSS) and promising:
the spiral wave grows by extracting energy and angular momentum from the
disk, and transferring them to the Rossby vortex at the corotation
radius; the vortex twists the footpoints of magnetic field lines
threading the disk, and this torsion will in turn be propagated as
Alfven waves to the corona of the disk, where it might energize a jet or
an outflow.

As the CSS or the $\beta$-plane approximation, our model can retain its
essential simplicity only at the cost of discarding the gradients of
equilibrium quantities such as the density or magnetic field.  It is
thus important to remember the more complete result obtained in the
above-mentioned works : they have shown that the corotation resonance
depends in fact on the gradient of $W/\rho$ (the specific vorticity) for 
waves driven by pressure or self-gravity, or $W\rho/B_{0}^2$ (where 
$B_{0}$ is the equilibrium magnetic field) for waves driven by magnetic 
stresses.
\section{Rossby waves with differential rotation}
\label{sec:rossby}
These applications of the coupling between spiral density waves and
Rossby waves must be discussed in the specific context associated with a
given disk model, and we defer this to separate works.  In particular we
will make use of the fact that these equations can straightforwardly be
extended to include vertical motions and the vertical gradient of
equilibrium quantities, in order to study the vertical structure of the
waves.

Here, in order to illustrate the use of the Rossby Shearing Sheet, we
limit ourselves to present an exact solution for the propagation of
Rossby waves in a differentially rotating disk, in the incompressible
limit.  This result is new, although it might have been obtained much
more simply from the conservation of specific vorticity.  We believe
that, given the simplicity of the system of equations
(\ref{eq:fourlinD}-\ref{eq:fourlinh}), it can be applied
straightforwardly to a planetary atmosphere with zonal circulation.

In the incompressible limit, ($D=h=0$) equation (\ref{eq:fourlinR}) has
the exact solution:
\begin{equation}
    R=R_{0}\exp\left(i\frac{\alpha}{2Am}\arctan\frac{k}{m}\right)
    \label{eq:rossk}
\end{equation}

Let us first consider this solution with the envelope function:
\begin{equation}
    F(k-2Amt)=Cte.\nonumber
\end{equation}
The inverse Fourier transform gives:
\begin{equation}
    R(s,t)=R_{0}\int_{-\infty}^{\infty}dk\exp\left[i\left(ks+\frac{\alpha}
    {2Am}\arctan\frac{k}{m}\right)\right]
    \label{eq:rossmode}
\end{equation}
which is independent of $t$ (\ie a normal mode solution if, by the choice 
of the frequency $\omega=m\Omega_{0}$, we manage to verify the appropriate 
boundary conditions). 

\begin{figure}
\psfig{file=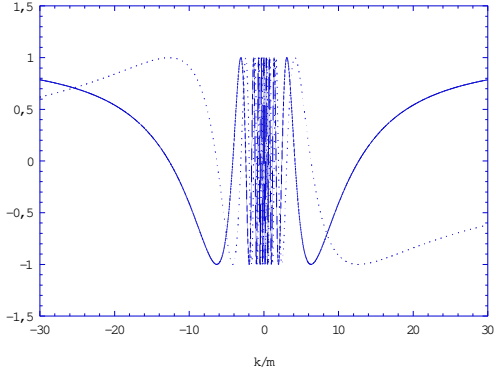,width=\columnwidth}
\caption{\label{fig:fig1} \scriptsize
Real (solid) and imaginary (dots) parts of the solution, equation 
(\ref{eq:rossk}) for very weak differential rotation ($\alpha/2Am=20$). 
The finite imaginary part at $k\rightarrow/\infty$ is responsible for 
the singular behavior of the wave as $t\rightarrow \infty$.
}\end{figure}

We will first discuss a case with weak differential rotation,
$A\ll\alpha$.  Figure (\ref{fig:fig1}) shows the integrand, in equation
(\ref{eq:rossk}), for $\alpha/2Am=20$.
The exponent in equation (\ref{eq:rossmode}) varies rapidly when $k\sim
m$.  The method of stationary phase tells us that at a given radius $s$
the main contribution to the integral in equation (\ref{eq:rossmode})
comes from the value of $k$ such that the exponent is stationary:
\begin{equation}
s+\frac{\alpha}{2A(k^2+m^2)}=0
\label{eq:kross}
\end{equation}

Using equation (\ref{eq:tom}) we see that this gives a local wavenumber
$k(s)$ such that it verifies the dispersion relation of Rossby waves,
Doppler-shifted by the equilibrium rotation:
\begin{equation}
    \omega-m\Omega(s)=\frac{m\alpha}{k^2+m^2}
\end{equation}
with $\omega=m\Omega_{0}$.  The wave propagates between $s=0$ 
(corotation, at large $k$) and a turning point ($k=0$) at:
\[s_{max}=-\frac{\alpha}{2Am^2}\]

Let us now consider the solution with 
\[F=\delta(k-2Amt):\]
the $\delta$ function has the effect that the full solution $R(k,t)$
explores the separated one, equation(\ref{eq:rossk}), while $k$ varies
from tightly wound leading ($k\rightarrow -\infty$) to tightly wound
trailing ($k\rightarrow \infty$) because of the shear.  The inverse
Fourier transform gives:
%\newpage
%
\begin{eqnarray}
    R(s,t)&=&R_{0}\int_{-\infty}^{\infty}dke^{iks}\delta(k-2Amt)\nonumber\\
    &&\hskip 3 cm \exp\left(i\frac{\alpha}
    {2Am}\arctan\frac{k}{m}\right)\nonumber\\
    &=& R_{0}\exp\left[i\left(ks+\frac{\alpha}
    {2Am}\arctan\frac{k}{m}\right)\right]
    \label{eq:rosstime}
\end{eqnarray}
where now $k=2Amt$.  Again in the limit where differential rotation is weak, 
$A\ll\alpha$, the stationary phase tells us that at a given $s$ the 
perturbation is strong only when $k$ has the value given in equation 
(\ref{eq:kross}), {\it and at the corresponding time}: we now have a 
description of a travelling Rossby wave, propagating from tightly 
leading ($k$ large and negative) at corotation 
at $t=-\infty$, to the turning point (at $t=0$) and back to corotation 
as tightly trailing ($k$ large and positive). 

 A qualitatively new and important result is obtained by considering the
 evolution of the wave at corotation: as $k$ varies the arctangent in
 equation (\ref{eq:rosstime}) varies from $-\pi/2$ to $\pi/2$, so that
 the exponent varies between $\pm {\pi\alpha}/{4Am}$.

{\em This means that at corotation the Rossby 
wave oscillates only 
\begin{equation}
    n_{oscillations}=\frac{\alpha}{4Am}
    \label{eq:nosc}
\end{equation}
times before it gets sheared away by differential rotation.  It always
evolves, as $t\rightarrow \infty$, to a singular vortex (a jump in
$v_{\vartheta}$ at $s=0$).  This final evolution will occur even for
arbitrarily weak differential rotation.}

At $s\neq 0$ the solution still oscillates with time because of the $ks$
term in the exponent: this is only due to the fact that a feature which
is stationary at $s=0$ is seen Doppler-shifted by differential rotation
by an observer at a different radius.  These results are still valid
(since the solution, equation (\ref{eq:rossk}) is exact) when
differential rotation is not weak, as in a keplerian disk, but their WKB
interpretation in terms of propagating waves becomes blurred.

At this stage we have thus described, in an equivalent (by the method of
stationary phase) to a WKB model, the propagation of a single Rossby
wave in a disk with differential rotation.

In the normal mode solution ($F=Cte$), at $s=0$, the initial and final
values of the integrand in equation (\ref{eq:rossmode}) are different,
also giving by inverse Fourier transform a singularity at
corotation. In the standing pattern obtained with this choice of $F$, 
the Rossby wave becomes a stationary vortex.

In general the envelope function $F$ is given by the initial conditions,
so that if they are regular initially (\eg describing a localized wave
packet) the Fourier transform will be well-behaved at infinity.  The
evolution will nevertheless lead to the same conclusion.

In the more general case where $A$ is not small, this evolution will be
very rapid: in a keplerian disk ($A=\alpha$) the wave will describe only
a fraction $1/4m$ of a cycle before it degenerates!  This is the reason
why Rossby waves cannot be seen in disks with a realistic vorticity
gradient.  It is also the reason why the corotation resonance is more
efficient at low $m$, so that the CSS can be considered as a convenient
approximation for $m\gg 1$.
\section{Evolution of vortices}
\label{sec:vortex}
\begin{figure}
\psfig{file=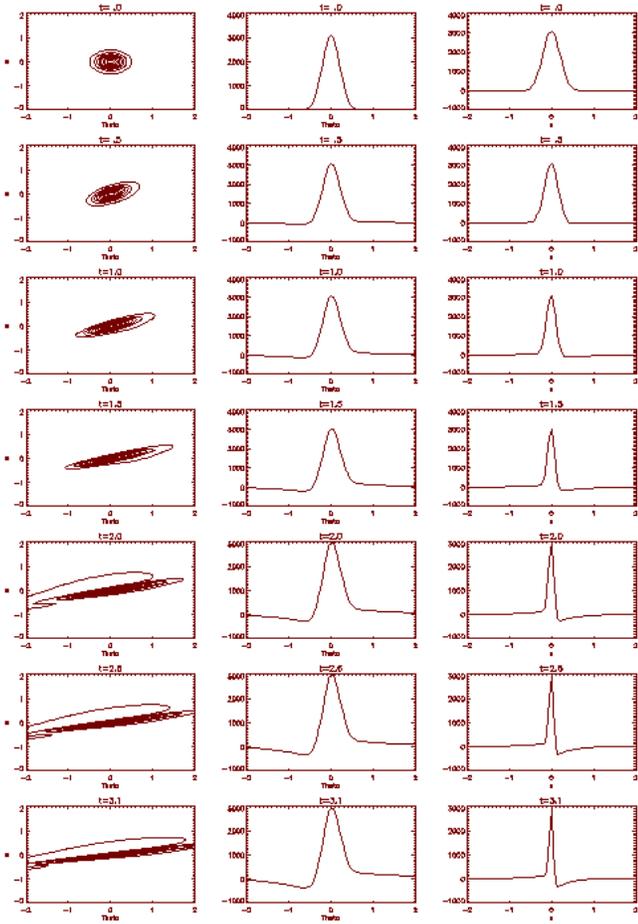,width=\columnwidth} 
\caption{\label{fig:figvort}
\scriptsize Evolution of a vortex in a differentially rotating disk. 
Contour plots, as well as cross sections at $\vartheta=0$ and $s=0$, are
shown as time varies from $0$ to $\pi$, {\em i.e.} during a half
rotation period. 
}\end{figure}
\begin{figure}
\psfig{file=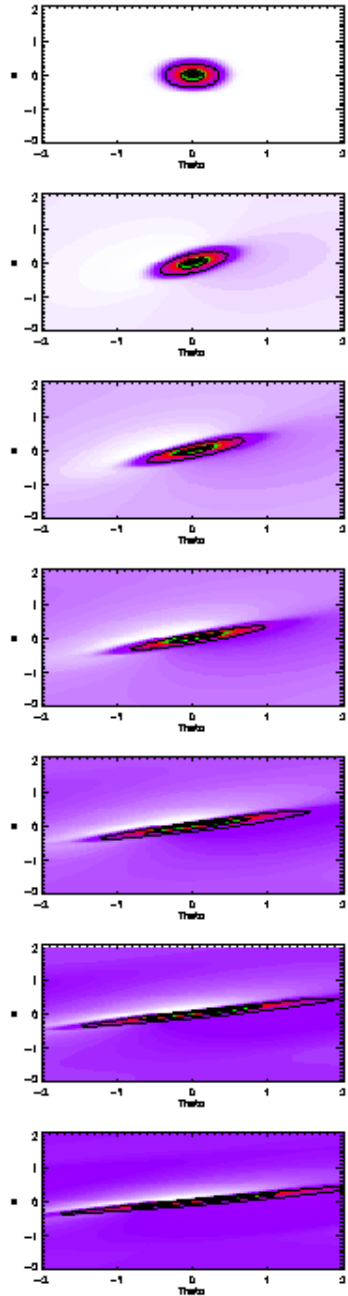,width=5cm}
\caption{\label{fig:imvort} \scriptsize
Same as figure \ref{fig:figvort}, with color-coded contours showing more 
details.
}\end{figure}
Recent numerical studies (Bracco {\em et al.}, 1998; Godon and Livio,
1999, 2000; Davis {\em et al.}, 2000) have addressed the fate of vortices in
protoplanetary disks.  This was prompted by the work of Barge and
Sommeria (1995), who had shown that a vortex would favor the formation
of planets, by capture and coagulation of dust particles.  These studies
show that, as expected, vortices are rapidly sheared apart by
differential rotation unless they are strong enough that non-linear
effects allow them to survive for a significant time.  This is obtained
by Godon and Livio (1999, 2000) as embedded cyclonic and anticyclonic
vortices.  Davis {\em et al.} obtain longer lived anticyclonic vortices. 
The difference probably lies in the initial conditions given.  In both
cases this requires such strong amplitudes (locally reversing the
rotation velocity gradient in the work of Davis {\em et al.}) that the
mechanism which could create such vortices (supersonic if they are
larger than the disk thickness) remains mysterious.

Here we will give a simple description of the linearized evolution of a
vortex.  We start from an initial gaussian vorticity perturbation, whose
Fourier transform is:
\begin{equation}
    R^0(m,k)=e^{-(m^2/m_{0}^2+k^2/k_{0}^2)}
    \label{eq:R0}
\end{equation}
with $m_{0}=k_{0}=7$ for the numerical solution displayed.  Using the
previous results, we find that the envelope function for each Fourier 
component is given by:
\begin{equation}
    F_{m}(k)=R^0(m,k)\exp\left[-i\frac{\alpha}
    {2Am}\arctan\frac{k}{m}\right]
    \label{eq:envelope}
\end{equation}
giving the time evolution of the vorticity (after a convenient change of
variables, $k\rightarrow k+2Amt$, in the inverse transform):
\begin{eqnarray}
    &R&(s,\vartheta,t)=
    \sum_{m}e^{im(\vartheta+2Ats)}e^{-m^2/m_{0}^2}
    \int_{-\infty}^{+\infty}  
    dke^{iks}e^{-k^2/k_{0}^2}\nonumber\\
    &&\ \exp\left({i\frac{\alpha}{2Am}
    \left[\arctan(\frac{k}{m}+2At)-\arctan(\frac{k}{m})\right]}\right)
\end{eqnarray}
which is easily computed (as it would be with any other choice of
initial condition).  The result is displayed in figures
\ref{fig:figvort} and \ref{fig:imvort}.  It shows that, as expected, the
vortex is rapidly sheared away by differential rotation, reducing to the
vorticity sheets described by Davis {\em et al.} (2000).  It applies
(since this is linear theory) to positive or negative vorticity, {\i.e.}
cyclonic or anticyclonic vortices, and corresponds to the evolution
observed for all but the highest amplitude vortices studied in
non-linear simulations. An obvious extension would be to include 
compressibility, {\em i.e.} return to the full third-order system of 
equations, and solve it numerically for the equivalent of the 
incompressible solution, equation (\ref{eq:rossk}).
\section{Discussion}
\label{sec:discuss}
We have derived, from the equations describing a differentially rotating
disk, a model (which we call the Rossby Shearing Sheet) which retains
the intrinsic simplicity of the conventional Shearing Sheet (obtained
with only differential rotation) or the $\beta$-plane approximation
(obtained with only differential vorticity) while including both
gradients.  The model results in a third order system of ordinary
differential equations, rather than second order as in the CSS or first
order in the $\beta$-plane approximation.  This means that, besides
inward- and outward-propagating spiral density waves, it incorporates
the physics of Rossby waves.

In a first application, we have used incompressibility to separate the
propagation of Rossby waves; we have shown that, for any ratio
$A/\alpha$ between differential rotation and differential vorticity, the
wave always collapses after a finite time to a singular vortex (a jump
in $v_{\vartheta}$ at its corotation radius).  In a realistic disk, this
evolution is so fast that it occurs before the wave has completed even
one cycle of oscillation.  In a planetary atmosphere, on the other hand,
zonal circulation should result in a lower ratio $A/\alpha$ and a slower
evolution of the wave, but still in its collapse to a singular vortex in
a finite time.  

We have also described the linear evolution of a vortex in a keplerian
disk, showing (as observed in non-linear simulations) that it reduces to
a thin vorticity sheet in a very short time.  This confirms that, unless
a very strong non-linear mechanism can be found to maintain a vortex
against shear, there is little hope of forming planets in this manner.

Lifting the constraint of incompressibility, we find that the density
waves and Rossby waves are no more independent: this means that they are
coupled by differential rotation in the disk.  Waves loose their
identity (the solutions are no more separated), so that a density wave
generates a Rossby wave, and reciprocally.  This process, in which
energy and angular momentum are exchanged leading to growth or damping
of the waves, is known as {\em corotation resonance} (but was not
analyzed in terms of coupling to the Rossby wave) in studies of the
spiral instability, with or without the long-range action of
self-gravity or magnetic stresses.  In the latter case it has recently
been found to make an important contribution, leading to what we have
called an Accretion-Ejection instability.  In recent works, Lovelace
{\em et al.} (1999) and Li {\em et al.} (2000) have also found an
instability of unmagnetized disks, when the specific vorticity profile
has an extremum.

Actually our result already allows us to give a physical interpretation
for the sign of the corotation resonance: we found that a Rossby wave
can propagate only between corotation and a radius
$s_{max}=-\alpha/2Am^2$, {\i.e.} inside corotation if $\alpha$ and $A$
have the same sign: since a wave propagating inside corotation has
negative energy and angular momentum (see {\em e.g.} Collett and
Lynden-Bell, 1987) we conclude that a spiral wave propagating inside
corotation will be damped by exciting a Rossby wave.  On the other hand
we have mentioned that in fact the relevant gradient for Rossby waves is
that of specific vorticity, $W/\Sigma$, or of $W\Sigma/B^2$ in a
magnetized disk.  Thus if these gradients change sign, they will also
change the sign of $s_{max}$ so that the Rossby wave will now propagate
{\em beyond} corotation, and will have a positive energy.  This means
that exciting it will {\em amplify} a spiral wave.

Future work will be dedicated to using this model for a deeper analysis
of the coupling between density and Rossby waves, in accretion disks
and in planetary atmospheres.

%%%%%%%%%%%%%%%%%%%%%%


\begin{thebibliography}{}
    
\bibitem{Barge}Barge, P. \& Sommeria, J., 1995, A\&A {\bf 295}, L1
    
\bibitem{Binney}
Binney, J. and Tremaine, S., 1987, {\it Galactic Dynamics}, 
Princeton University Press

\bibitem{Bracco}Bracco, A., Chavanis, P.H. \&Provenzale, A., 1998, Phys. 
Fluids {\bf 11}, 2280

\bibitem{Collett}Collett, J.L. \& Lynden-Bell, D., 1987, MNRAS {\bf 
224}, 489

\bibitem{Davis}Davis, S.S., Sheehan, D.P., and Cuzzi, J.N., 2000, ApJ 
{\bf 545}, 494

\bibitem{Godon}Godon, P. \& Livio, M., 1999, ApJ {\bf 523}, 350

\bibitem{Godon2}Godon, P. \& Livio, M., 2000, ApJ {\bf 537}, 396

\bibitem{Li}
Li, H. Finn, J. M. Lovelace, R. V. E. and Colgate, S. A., 2000, ApJ 
{\bf 533}, 1023L

\bibitem{Lin}
Lin,C.C., and Thurstans , R.P.: 1984, {\it  in } Proc. of a course on Plasma
Astrophysics, Varenna, Italy, ESA S.P. {\bf 207} (Noordwijk, Holland), 121

\bibitem{Lovelace}
Lovelace, R. V. E., Li, H., Colgate, S. A. and Nelson, A. F., 1999, ApJ 
{\bf 513}, 805L

\bibitem{Narayan}
Narayan, R., Goldreich, P. and Goodman, J., 1987, MNRAS {\bf 228}, 1

\bibitem{Pannatoni}
Pannatoni, R.F., 1983, Geophys. Ap. Fluid Dyn. {\bf 24}, 165

\bibitem{Papaloizou}
Papaloizou, J.C.B. and Pringle, J.E., 1984, MNRAS {\bf 208}, 721

\bibitem{Papaloizou2}
Papaloizou, J.C.B. and Pringle, J.E., 1985, MNRAS {\bf 213}, 799

\bibitem{Papaloizou3}
Papaloizou, J.C.B. and Lin, D.N.C., 1989, ApJ. {\bf 344}, 645

\bibitem{THSP}
Tagger, M., Henriksen, R.N., Sygnet, J.F. and Pellat, R., 1990, ApJ {\bf
353}, 654

\bibitem{Tagger}
Tagger, M., and Pellat, R., 1999, A\& A {\bf 349}, 1003

\end{thebibliography}
\end{document}